\documentclass[12pt]{article}

\usepackage{scicite}

\usepackage{times}

\usepackage{booktabs}
\usepackage{url}
\usepackage{xcolor}
\usepackage[colorlinks = true,
            linkcolor = blue,
            urlcolor  = blue,
            citecolor = blue,
            anchorcolor = blue]{hyperref}

\usepackage{graphicx}
\usepackage{amsmath}
\usepackage{amsfonts}
\usepackage{bm}

\usepackage{amsthm}
\usepackage{diagbox}
\usepackage{multirow}
\usepackage[ruled]{algorithm2e}
\usepackage{color}
\newcommand{\red}[1]{\textcolor{black}{#1}}
\usepackage{soul}

\newenvironment{sciabstract}{%
\begin{quote} \bf}
{\end{quote}}

\title{X-ray Dissectography Improves Lung Nodule Detection}

\author
{Chuang Niu,$^{1}$ Giridhar Dasegowda,$^{2}$ Pingkun Yan,$^{1}$ \\
Mannudeep K. Kalra,$^{2}$ Ge Wang$^{1}$\\
\\
\normalsize{$^{1}$AI-based X-ray Imaging System (AXIS) Lab}\\
\normalsize{Department of Biomedical Engineering, Rensselaer Polytechnic Institute,}\\
\normalsize{110 8th Street, Troy, New York 12180, USA}\\
\normalsize{$^{2}$Department of Radiology}\\
\normalsize{Massachusetts General Hospital, Harvard Medical School}\\
\normalsize{White 270-E, 55 Fruit St, Boston, MA 02114, USA}\\
\\
}

\date{}

\begin{document} 


\baselineskip24pt


\maketitle


\begin{sciabstract}
\textcolor{black}{Although radiographs are the most frequently used worldwide due to their cost-effectiveness and widespread accessibility, the structural superposition along the x-ray paths often renders suspicious or concerning lung nodules difficult to detect.}
In this study, we apply ``X-ray dissectography" to dissect lungs digitally from a few radiographic projections, \textcolor{black}{suppress} the interference of irrelevant structures \textcolor{black}{and improve} lung nodule detectability.
For this purpose, a collaborative detection network is designed to localize lung nodules in 2D dissected projections and 3D physical space.
Our experimental results show that our approach can significantly improve the average precision by 20+\% in comparison with the common baseline \textcolor{black}{that detects lung nodules from original projections using a popular detection network.}
Potentially, this approach could help re-design the current X-ray imaging protocols and workflows and improve the diagnostic performance of chest radiographs in lung diseases.

\end{sciabstract}


\section{Introduction}

X-ray imaging is the first and still most popular medical imaging modality, which is performed by various kinds of systems.
\textcolor{black}{In the low-end, x-ray radiography takes a two-dimensional (2D) projective image through a patient, which is called a radiogram or radiograph.
In the high end, many x-ray projections are first collected and then reconstructed into computed tomographic (CT) images transversely or volumetrically.}
Before performing a typical CT scan, 2D scout views, \textcolor{black}{also konwn as topograms or planning radiographs,} are initially obtained for prescribing CT slices and modulating x-ray tube current for dose reduction. 
\textcolor{black}{Radiologists have found that planning radiogaphs carry significant diagnostic information \cite{nazir2013missed, ctscout, daffner2015reviewing, ichikawa2021deep},
although they have limited information compared to radiography and CT}.
These X-ray imaging modes have their strengths and weaknesses.
X-ray projections, either \textcolor{black}{radiographs or planning radiographs}, have the advantages of \textcolor{black}{fast scan speed and low radiation dose}, but they have multiple organs and tissues superimposed along x-ray paths in projections, which compromise the diagnostic performance.
On the other hand, X-ray CT unravels structures overlapped in the projection domain into tomographic images in the generic image space but a CT scanner is rather complicated, costly, and takes a \red{comparatively} higher radiation dose.

Lung cancer is the most commonly diagnosed cancer (11.6\% of the total cases) and the leading cause of cancer death (18.4\% of the total cancer deaths) worldwide \cite{lungcancer}.
X-ray radiography and CT are two common means to image the \red{chest abnormalities}. While CT is more \red{accurate} but also more expensive, standard chest X-ray (CXR) screening is an important and much more accessible way for detection of several chest abnormalities across the world.
\red{Like CXRs, CT planning radiographs acquired to set up the CT exam can} 
provide substantial clinically relevant information
\red{not demonstrated on the reconstructed axial images for lung cancer detection}\cite{nazir2013missed}.
Although \red{these 2D X-ray projections,} \red{either CXRs or planning radiographs} are informative \red{for the lung cancer detection, as above mentioned}, the structural superimposition in 2D projections makes lung nodules rather challenging to be  detected. The nodule visibility depends on various factors, such as size, location, projection angle, attenuation, and intervening features along x-ray paths.
\red{Such variations are compounded by substantial variations between radiologists on reporting of nodules on CXRs.}

Computer aided diagnosis (CAD) systems were developed for lung nodule detection.
Recent studies \red{demonstrate} that artificial intelligence (AI) algorithms as the second \red{or concurrent} reader help radiologists in detection of lung cancers on chest radiographs \cite{yoo2021ai}.
Most of the existing detection algorithms seem having limited room for improving the detection performance, unless better image quality and more prior knowledge are provided.
For example, it has been demonstrated that suppressing ribs in CXR images can help radiologists in the lung nodule detection task \cite{freedman2011lung}.
On the other hand, some methods learned to map 2D \red{projection radiographs} to 3D CT volumes and achieved remarkable results, they cannot reconstruct structures accurately and reliably \cite{Ying_2019_CVPR, shen2019patient, shen2021geometryinformed}, and their clinical utilities have not been demonstrated so far.
All the above methods depend on the GAN framework and/or unpaired learning for 2D/3D image generation. A major potential problem is that GAN-based models tend to generate fake structures, which is undesirable in the medical imaging field.

In this study, we develop x-ray dissectography (XDT) to digitally dissect a target organ/tissue and a collaborative network to synergistically detect lung nodules in both 2D and 3D spaces from several 2D X-ray projections.
Different from conventional rib suppression that only removes ribs on projection images, XDT is to extract isolated lungs removing all other components via deep learning.
However, it is not feasible to obtain ground-truth radiographs of a dissected organ of interest in a living patient for network training.
Naturally, we leverage widely available CT volumes in the training stage.
In other words, a number of previously obtained 3D CT scans can be used as prior to improve 2D radiograph or planning radiographs based lung nodule detection using a deep neural network after being trained in the supervised mode.
To obtain X-ray projections of a target organ without surrounding tissues, we can segment the organ out of the corresponding CT volume and then generate projections through the segmented organ according to the imaging system parameters.
Since a planning radiographs is generated as an X-ray radiograph by linearly scanning a fan-beam of x-rays before performing a CT scan, real and target planning radiographs can be well aligned for a given CT volume as a training pair.
Hence, we mainly focus on improving planning radiographs quality and detecting lung nodules from the enhanced planning radiographs. In principle, this approach can be extended to radiography after moderate adaptions.
As different projections contain complementary information, we design an XDT network to synergistically extract lungs from more than one projection view.
Indeed, multiple projection views benefit lung nodule detection. A nodule missed in one projection view may be clearly observed in another view.
On the other hand, lung nodules can be localized in 3D from more than one views. Currently, there is no method that collaboratively detects lung nodules and predicts their 3D locations.
To predict 2D and 3D locations simultaneously and collaboratively, we first extract 2D features of all projections and back-project these 2D features to form 3D features, which are respectively followed by 2D and 3D detectors in an end-to-end fashion. Finally, a collaborative matching algorithm is used to integrate all 2D and 3D predictions by enhancing from each other.

Our experimental results have shown that XDT separates the lungs well with faithful texture and structures.
Promisingly, taking three projection views as inputs, our proposed method improves average precision (AP) of lung nodule detection by 20+\% (34.7\% v.s. 62.1\%) in comparison with the common baseline method that detects lung nodules from a single 2D projection. Our deep learning code will be made publicly available.

\section{Method}

\begin{figure*}[htp]
    \centering
    \includegraphics[width=1\textwidth]{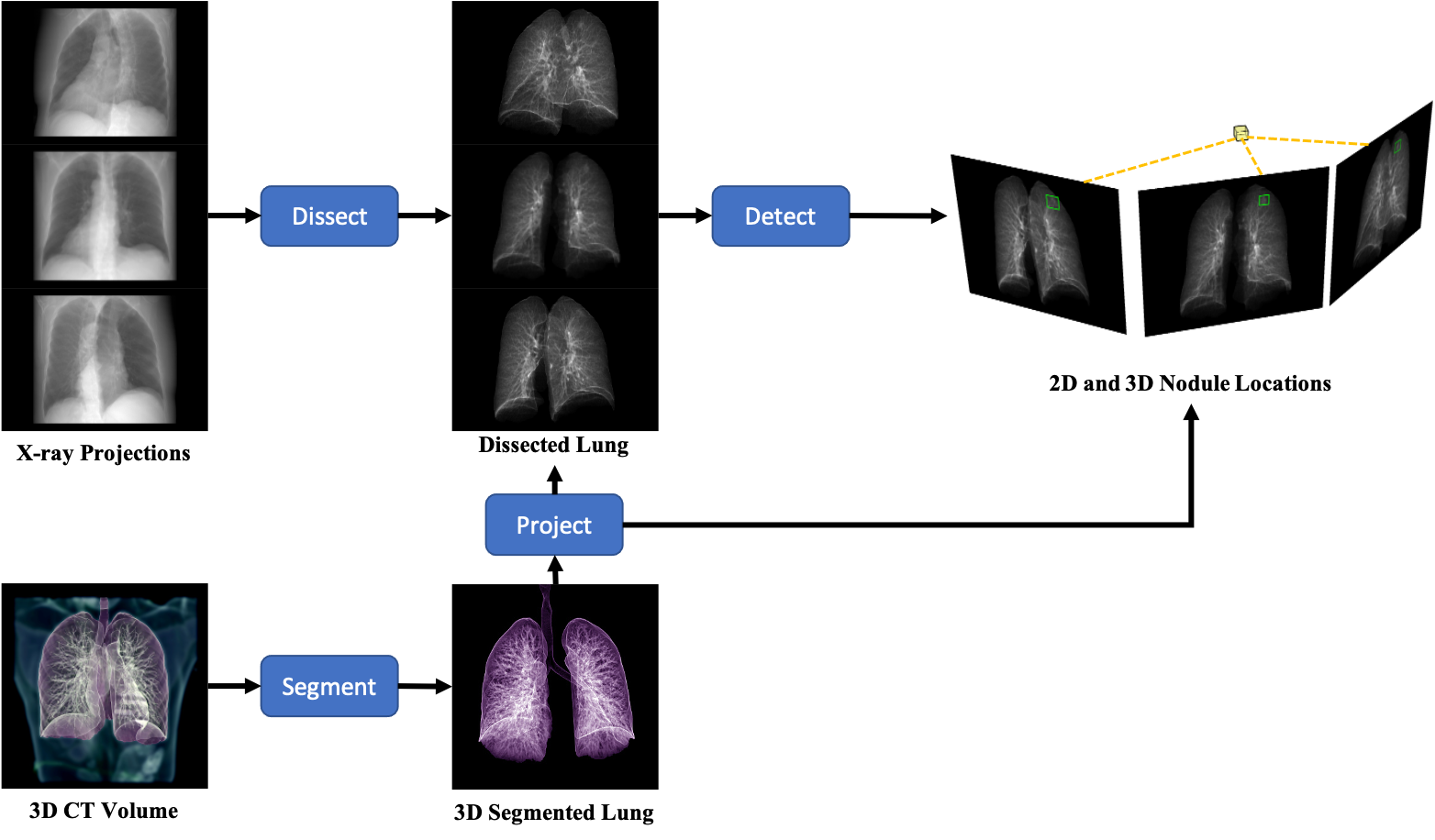}
    \caption{Dissectography based collaborative lung nodule detection.}
    \label{fig:framework}
\end{figure*}

We propose to computationally isolate lungs out of the check and collaboratively detect lung nodules from more than one 2D X-ray projections through a deep learning framework, as shown in \ref{fig:framework}.
To achieve this goal, an XDT network is designed to dissect lungs in more than one projections.
Then, a collaborative 2D-3D network is proposed to localize lung nodules in 2D and 3D spaces simultaneously.
In the training stage, 3D CT volumes are used as the prior knowledge to generate projections for dissection and detection according to the imaging geometry.
In the reference stage, only several 2D X-ray projections are required and forwarded to the trained XDT and collaborative detection networks, outputting the dissected lung projections along with 2D and 3D locations of detected nodules. In the following, we describe the XDT and collaborative networks in details.

\subsection{X-ray Dissectography Network}

\begin{figure*}[htp]
    \centering
    \includegraphics[width=1\textwidth]{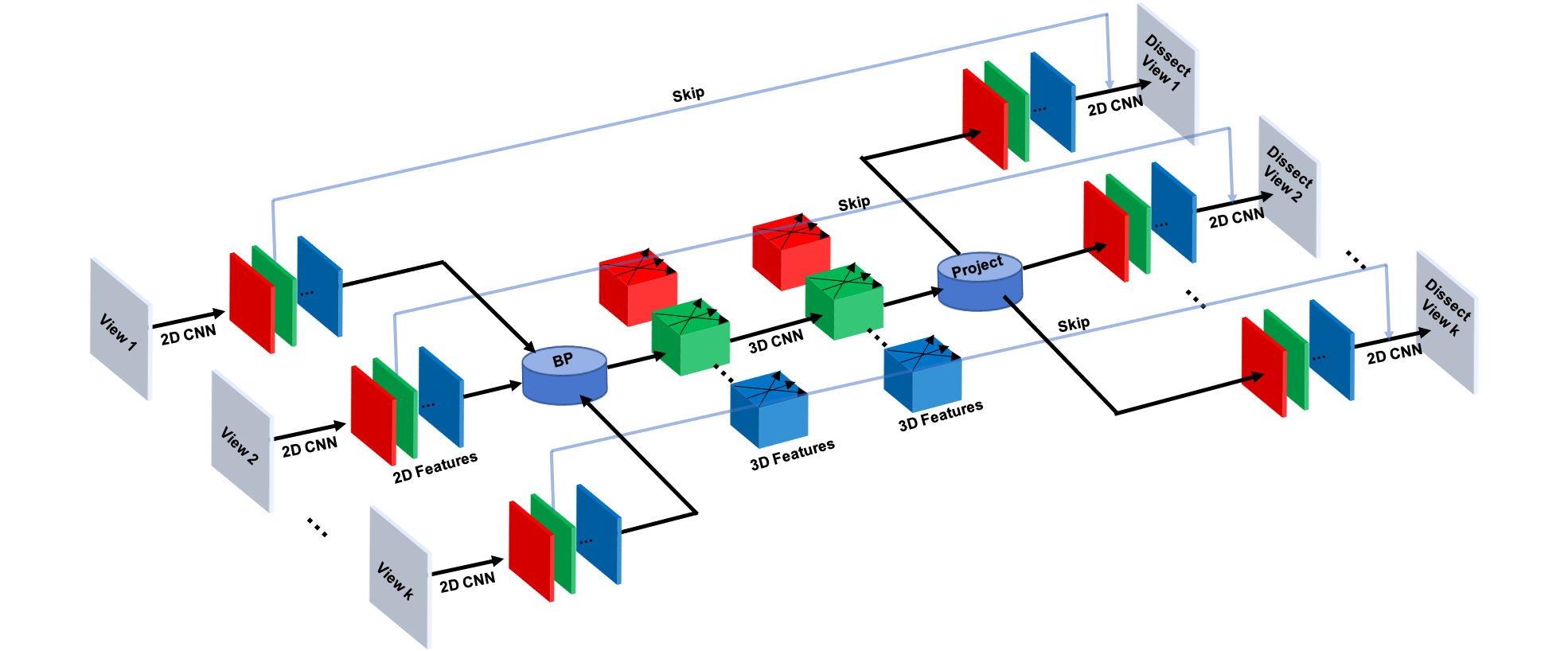}
    \caption{X-ray dissectography network.}
    \label{fig:xdt}
\end{figure*}

Given several 2D X-ray projections, the XDT network outputs isolated lungs in each of the projections.
As shown in Fig. \ref{fig:xdt}, XDT consists of five components: a 2D feature encoder, a feature back-projection layer, a 3D feature transformation layer, a feature projection layer, and a 2D feature decoder.
Specifically, a set of 2D X-ray projections are respectively forwarded to the 2D convolutional neural networks (CNN) with shared weights to extract 2D features.
As 2D CNN features of a specific channel (e.g., the red channel) can be regarded as the projections of a certain type of 3D features from different views, we independently reconstruct a 3D volume for each and every channel through a 3D reconstruction layer, which is implemented as the back-projection (BP) operation using the same imaging parameters. The equations for BP and projection can be found in \cite{zeng2010medical}.
It is worth mentioning that the traditional 3D reconstruction requires a large memory and a long runtime so that it is impractical to be integrated into the neural network for end-to-end training.
In this study, this reconstruction layer takes 2D CNN features of a much lower spatial resolution (16$\times$ 16 less in 2D and 16$\times$16$\times$16 in 3D) and a much less number of projections (several instead of hundreds or thousands of views).
In this way, the 2D features can be exactly aligned into 3D features in consistence with the physical imaging process.
Then, a 3D CNN layer further analyzes 3D features in multiple channels.
To leverage complementary information from different views for this regression task, we project the 3D CNN features of all views back to the 2D feature space via a projection layer.
The projection layer is the dual operation of the BP layer.
Finally, symmetric 2D feature decoder with skip connections from the encoder component is implemented to regress the final dissection results.
Here the $L1$ loss function is defined as follows:
\begin{equation}
    L_{XDT} = \frac{1}{N\times K} \sum_{i=1}^N \sum_{k=1}^K ||f_{XDT}(\bm{x}_{ik}, \bm{\theta}_{XDT}) - \bm{y}_{ik}||_1
\end{equation}
where $N$ and $K$ are respectively the numbers of samples and projection views, $f(\cdot, \bm{\theta}_{XDT})$ is the network function with a vector of parameters $\bm{\theta}_{XDT}$ to optimize, and $\bm{y}_{ik}$ denotes the target projection of the isolated lungs.

\subsection{Collaborative Detection Network}

\begin{figure*}[htp]
    \centering
    \includegraphics[width=1\textwidth]{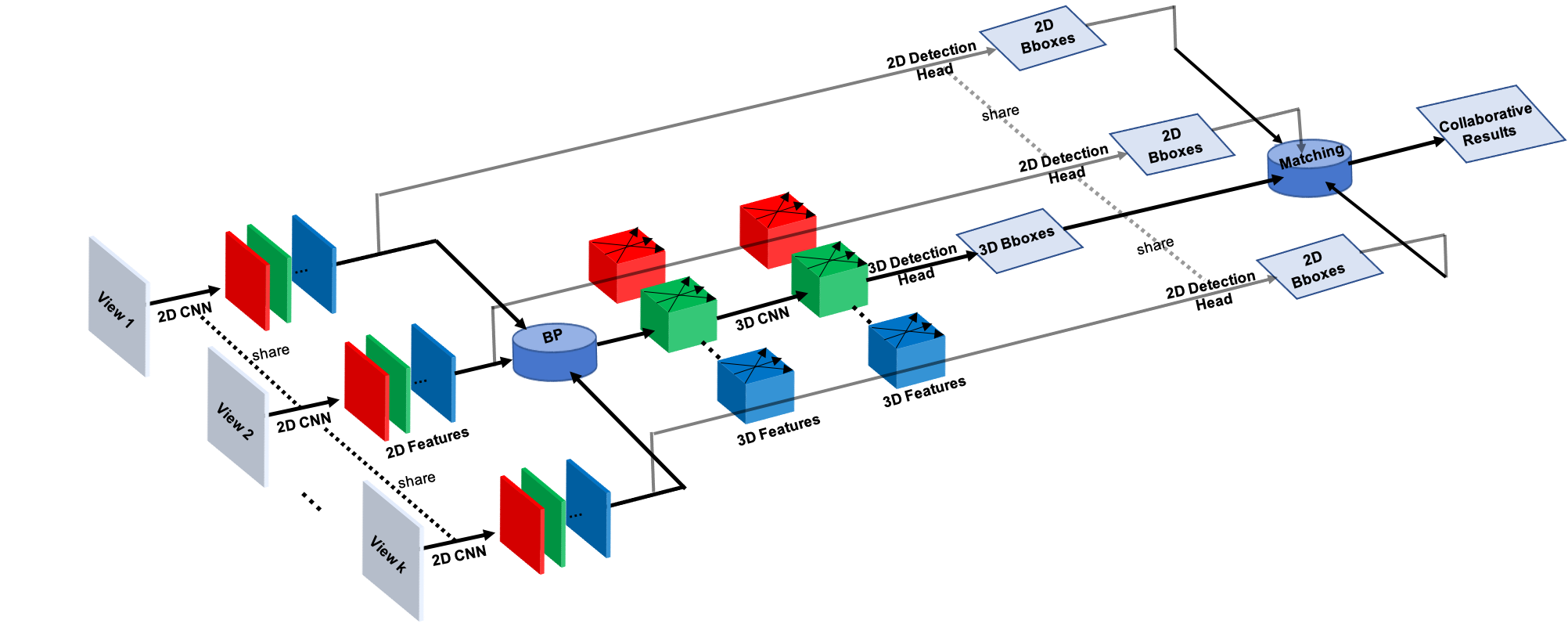}
    \caption{Collaborative detection network.}
    \label{fig:detection}
\end{figure*}

Different projection views not only provide complementary information of lung nodules but also help determine their 3D locations.
Typically, two X-ray projections are taken, one from the front and the other from the lateral side of the body.
If the nodule can be accurately detected from these two orthogonal views, the 3D location can be exactly determined.
However, it is difficult to detect nodules from the lateral view where more organs and tissues are superimposed.
In clinical scenarios, additional views may be required for better evaluation.
The existing AI algorithms for lung nodule detection can only process single views separately without providing 3D locations.

In this section, we propose a dedicated network architecture to collaboratively detect lung nodules from more than one projections and output their locations in both 2D and 3D.
Over past years, many object detection networks \cite{ren2015faster, Redmon_2016_CVPR, Tian_2019_ICCV} were proposed for various applications \cite{nndetect, 8410588, 8667713}.
Here we adopted the most popular Faster RCNN \cite{ren2015faster} detector as our baseline model.
Nevertheless, we do not see any obstacles to implement other object detectors into our collaborative detection architecture.
As shown in Fig. \ref{fig:detection}, the proposed model consists of five components: a 2D feature encoder, a 2D object detector, a back-projection layer followed by a 3D CNN, a 3D object detector, and a collaborative matching algorithm.
Specifically, a series of projections are forwarded to 2D feature encoders with shared weights to extract their 2D CNN features.
In the upper branches, these 2D features of different views are input to the 2D object detectors with shared weights for separate detection of nodules in different projections.
In the lower branches, these 2D features are back-projected into the 3D space underlying the physical imaging process via the back-projection layer, which is the same as that introduced in the XDT network.
The obtained multi-channel 3D features are further integrated through the subsequent 3D CNN.
Taking 3D features as inputs, the 3D object detector is designed to predict and localize lung nodules in the physical 3D space.
Finally, given these predicted 2D and 3D nodules, the collaborative matching algorithm produces the final 2D and 3D detection results systematically.
In the training stage, 2D and 3D detectors are simultaneously optimized. The collaborative matching algorithm contain neither parameters to optimize nor hyper-parameters to set, and can be directly applied in the inference stage.
Let us now present in detail the 2D and 3D detectors as well as the collaborative matching algorithm.

\subsubsection{2D and 3D Object Detectors}
For 2D detection, the two-stage Faster RCNN object detector is adopted. In the first stage, a region proposal network (RPN) generates a set of candidate bounding-boxes (BBox) that may contain objects of interest.
In the second stage, given the candidate BBoxes and the 2D CNN features, the region of interest (RoI) align layer is followed by a classification head and a box regression head to predict the object class and refine the BBoxes respectively. More details about each component can be found in \cite{ren2015faster}.
The 3D object detector is extended from the 2D object detector by modifying each 2D component to the 3D counterpart cosisting of the 3D anchors, 3D RoI align layer, 3D CNN classification and regression heads, and corresponding loss functions.
The final loss function for optimizing the whole collaborative detection network is defined as:
\begin{equation}
    \begin{split}
        L_{Det} =& \sum_{k=1}^{K} (L_{BCE}(p_{k}^{2Dr}, p^{2D*}_k) + L_{2Dloc}(t_{k}^{2Dr}, t^{2D*}_k) + L_{BCE}(p_{k}^{2D}, p^{2D*}_k) + L_{2Dloc}(t_{k}^{2D}, t^{2D*}_k)) \\
                 & + L_{BCE}(p^{3D}, p^{3D*}) + L_{3Dloc}(t^{3D}, t^{3D*}) 
    \end{split}
\end{equation}
where $L_{BCE}$ denotes the binary cross entropy for identifying if 2D and 3D BBoxes contain nodules or not, $L_{2Dloc}$ and $L_{3Dloc}$ are the BBox regression loss functions for localizing nodules in 2D and 3D spaces respectively, and the BBox regression loss is implemented as the smooth L1 function defined in \cite{Girshick_2015_ICCV}, $p_{k}^{2Dr}$ and $t^{2Dr}_k$ to compute the predicted classification scores and coordinates of 2D RPN BBoxes against the ground truth, $p_{k}^{2D}$ and $t^{2D}_k$ are the predicted classification scores and coordinates from the 2D RoI classification and regression heads, $k$ and $K$ represent projection index and the total number of projections, $p^{3D}$ and $t^{3D}$ are the predicted classification scores and coordinates of the 3D BBoxes, and $p^{2D*}_k$, $t^{2D*}_k$, $p^{3D*}$, and $t^{3D*}$ are the corresponding ground-truth.
Here we only implement one stage detector for predicting 3D locations, as we found that the second refinement stage cannot improve the performance due to the limited quality of the 3D features reconstructed from several views only.
We follow the parameterization strategy in \cite{Girshick_2014_CVPR} to encode both 2D and 3D coordinates. Since 3D coordinates include the 2D ones as a special case, here we only introduce the 3D parameterizations as follows:
\begin{equation}
    \begin{split}
        &t_x = (x - x_a) / w_a, t_y = (y - y_a) / h_a, t_z = (z - z_a) / d_a \\
        &t_w = log(w/w_a), t_h = log(h/h_a), t_d = log(d/d_a)
    \end{split}
\end{equation}
where $x, y, z$ are the center coordinates of a 3D BBox, $w, h, d$ are the width, height and depth (axial direction) of the 3D BBox, $x_a, y_a, z_a, w_a, h_a, d_a$ are the corresponding anchor box as defined in \cite{ren2015faster}.

\begin{algorithm}
\label{alg_1}
\caption{2D-3D Matching Algorithm}
\LinesNumbered
\KwIn{$\{\{b_{ik}^{2D}\}_{i=1}^{N_k}\}_{k=1}^K$, $\{b_i^{3D}\}_{i=1}^{N}$}
\KwOut{$\{(\{b_{ki}^{2Dout}\}_{k=1}^K, b_i^{3Dout})\}_{i=1}^{N_M}$}

Project $\{b_i^{3D}\}_{i=1}^{N}$ to $\{\{b_{ki}^{32D}\}_{k=1}^K\}_{n=1}^{N}$ \;

Calculate an IoU matrix $U = [u_{ik}=\max_{1\le j \le N_k}(IoU(b_{ki}^{32D}, b_{jk}^{2D}))]^{N\times K}$ \;
Calculate an index matrix $Q = [q_{ik}= -1$ or $\arg\max_{1\le j \le N_k}(IoU(b_{ki}^{32D}, b_{jk}^{2D}))]^{N\times K}$ \;
Compute a matched index matrix $Q^M = [q_{i:}]^{N_M \times K}$, $i = \arg\max_{e \in E} \sum_{k} u_{ek}/K$, $E = \{j: q_{jk} = q_{ik} \le 0, \forall j,k\}$ \;
Generate collaborative results $\{(\{b_{ki}^{2Dout}\}_{k=1}^K, b_i^{3Dout})\}_{i=1}^{N_M}$, $b_{ki}^{2Dout}=b_{kq_{ik}}^{2D}$ or $b_{kI_i}^{32D}$ \;
\end{algorithm}

\subsubsection{Collaborative Matching Algorithm}
Without any hyper-parameters, our collaborative matching algorithm can collaboratively integrate all 2D and 3D predictions.
The motivation is very heuristic. A nodule missed in one projection may be detected in another projection, and a strongly positive nodule in every projection will be easily detected in the integrated 3D space.
The overall procedure is summarized in Algorithm \ref{alg_1}.
Given the detected 2D BBoxes $\{\{b_{ik}^{2D}\}_{i=1}^{N_k}\}_{k=1}^K$ and 3D BBoxes $\{b_i^{3D}\}_{i=1}^{N}$, where $N_k$ is the number of 2D boxes in the $k$-th projection, and $N$ is the number of 3D BBoxes, the matching algorithm identifies which 2D and 3D boxes are from the same nodule. Thus, the nodules missed from a projection can be re-detected from other projections.

First, to match 3D with 2D BBoxes we project 3D BBox coordinates $(b_{x1}^{3D}, b_{y1}^{3D}, b_{z1}^{3D}, b_{x2}^{3D}, b_{y2}^{3D},$ $b_{z2}^{3D})$ to 2D BBox coordinates $(b_{x1}^{32D}, b_{z1}^{32D}, b_{x2}^{32D}, b_{z2}^{32D})$ according to the imaging parameters, where the 3D BBox is represented by the top-left-deepest and bottom-right-lowest vertexes and the the 2D BBox is represented by the top-left and bottom-right vertexes. Here we assume a simple case of a parallel beam projection geometry, and define a rotation transform $(b'_x, b'_y) = R(b_x, b_y, \theta)$ of 2D coordinates for a projection angle $\theta$ as follows:
\begin{equation}
    \begin{bmatrix}
    b'_x \\
    b'_y 
    \end{bmatrix} = 
    \begin{bmatrix}
    \cos{\theta} & \sin{\theta} \\
    -\sin{\theta} & \cos{\theta} \\

    \end{bmatrix}
    \begin{bmatrix}
    b_x - b_{xc} \\
    b_y - b_{yc}\\
    \end{bmatrix} + 
    \begin{bmatrix}
    b_{xc} \\
    b_{yc} \\
    \end{bmatrix}
\end{equation}
where $(b_{xc}, b_{yc})$ are the coordinates of the object center.
Then, the coordinates of rotated vertexes are computed as follows:
\begin{equation}
    b'_{x11} = R(b^{3D}_{x1}, b^{3D}_{y1}, \theta), b'_{x12} = R(b^{3D}_{x1}, b^{3D}_{y2}, \theta), b'_{x21} = R(b^{3D}_{x2}, b^{3D}_{y1}, \theta), b'_{x22} = R(b^{3D}_{x2}, b^{3D}_{y2}, \theta)
\end{equation}
The coordinates of 2D BBoxes are also computed:
\begin{equation}
    b_{x1}^{32D} = \min(b'_{x11},b'_{x12},b'_{x21},b'_{x22}), b_{x1}^{32D} = \max(b'_{x11},b'_{x12},b'_{x21},b'_{x22}), b_{z1}^{32D} = b_{z1}^{3D}, b_{z2}^{32D} = b_{z2}^{3D}
\end{equation}

Second, we compute the intersection over union (IoU) matrix $U = [u_{ik}]^{N\times K}$, where $u_{ik}=\max_{1\le j \le N_k}(IoU(b_{ki}^{32D}, b_{jk}^{2D}))$, $IoU(b_{ki}^{32D}, b_{jk}^{2D})$ means the IoU value between $b_{ki}^{32D}$ and $b_{jk}^{2D}$, and $u_{ik}$ is the maximum IoU value between the $i$-th 3D BBox and all detected 2D BBoxes in the $k$-th projection, 
and compute the corresponding index matrix $Q = [q_{ik}]^{N\times K}$ of matched 2D BBoxes to 3D BBoxes, where $q_{ik} =-1$ if $u_{ik} = 0$, which means that there is no 2D BBox in the $k$-th projection which is matched to the $i$-th 3D BBox, $q_{ik} = \arg\max_{1\le j \le N_k}(IoU(b_{ki}^{32D}, b_{jk}^{2D}))$ is the index of matched 2D BBox. 

Third, given the IoU and index matrices, we compute the final matched index matrix according to the matching principle that when multiple 3D boxes are matched with the same 2D BBox, only one 3D BBox having the maximum average IoU over all projection views is preserved, and the others are regarded as redundant and removed. Formally, the matched index matrix is denoted as $Q^M = [q_{i:}]^{N_M \times K}$, the preserved 3D index is denoted as $I=\{I_j=i\}^{N_M}_{j=1}$, where $i = \arg\max_{e \in E} \sum_{k} u_{ek}/K$, $E = \{j: q_{jk} = q_{ik}, \forall j,k\}$, and $N_M$ denotes the number of matched 3D BBoxes. 

Fourth, the collaborative 2D-3D detection results are generated according to the matched index matrix. When 2D BBoxes are missed in some projections views, they will be recovered as the projected 2D BBoxes of the corresponding 3D ones.
Finally, given the matched index matrix $Q^M = [q_{i:}]^{N_M \times K}$, the collaborative detection results are calculated as $\{(\{b_{ki}^{2Dout}\}_{k=1}^K, b_i^{3Dout})\}_{i=1}^{N_M}$, where $b_{ki}^{2Dout}=b_{kq_{ik}}^{2D}$ if $q_{ik} \ge 0$, else $b_{ki}^{2Dout}=b_{kI_i}^{32D}$.

In this matching algorithm, we use the 3D detection results as a bridge to explicitly enhance 2D detection results from different views. As the 3D detector is implemented as the one-stage RPN, initial results include many false positive BBoxes. This matching algorithm is then used to filter out false positive results effectively. Although 3D detection results include a majority of 2D results, there are still 2D detected BBoxes which do not match any 3D BBoxes. In the inference stage, we still need these unmatched 2D bboxes for final decision.

\section{Experimental Design and Results}

\subsection{Dataset}
In this study, the lung nodule detection dataset LIDC-IDRI \cite{armato2011lung} was used for training and testing our proposed method. LIDC-IDRI consists of 1,010 patient scans. Annotations in the LIDC-IDRI database were done by 4 experienced radiologists during a two-phase annotation process.
Each radiologist marked lesions they identified as non-nodule, nodule $<$ 3 mm, and nodules $\ge$ 3 mm.
Here we considered the nodules $\ge$ 20 mm accepted by at least 3 out of 4 radiologists.

\subsection{Data Preprocessing}
\begin{figure*}[htp]
    \centering
    \includegraphics[width=1\textwidth]{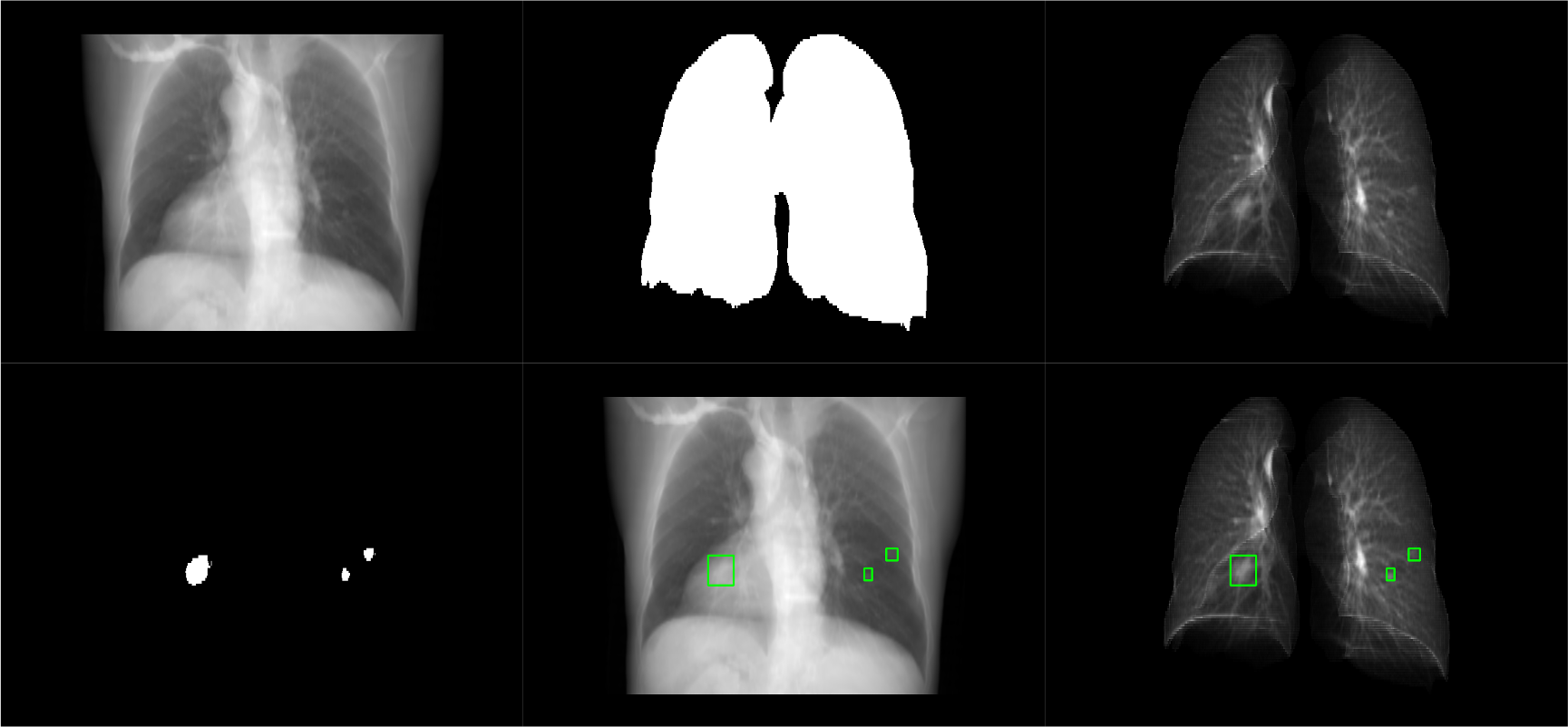}
    \caption{Projection examples. Top row from left to right: An original X-ray projection, lung mask, and lung projection; Bottom row from left to right: The lung nodule mask, lung nodule bounding-boxes on the original projection, and lung nodule bounding-boxes on the pure lung projection.}
    \label{fig:data}
\end{figure*}
We applied a trained lung segmentation model \cite{hofmanninger2020automatic} to generate a set of 3D lung segmentation masks.
Then, we merged these lung masks with both the corresponding lung masks provided in LUNA \cite{SETIO20171} and the lung nodule mask annotated in the original dataset to get the final lung mask, ensuring that the lung mask cover nodules and the lung region as completely as possible.
As the CT scans in this dataset have lower resolution in the axial direction than the other two directions, we first up-sampled with 
trilinear interpolation
the all CT scans so that the voxel size along the axial direction is the same as that in the XY plane, and also up-sampled the lung masks accordingly.
In this simulation, we assumed the parallel-beam projection geometry to generate the planning radiographs without lose of generality.
Specifically, we respectively projected the original 3D volume, 3D lung region volume, 3D lung mask, each 3D nodule mask to obtain original 2D projection containing all superimposed structures, 2D lung projection of the isolated lungs only, 2D lung mask and 2D bounding-box coordinates calculated from the 2D nodule mask, respectively.
The size of all 2D projections was set to $512 \times 736$. The example projections are shown in Fig. \ref{fig:data}.

\subsection{Dissection Results}

\begin{table}[htp]
  \renewcommand{\arraystretch}{1.5}
  \renewcommand\tabcolsep{30pt}
 \caption{Quantitative dissection metrics in terms of PSNR and SSIM.}
  \centering
  \begin{tabular}{c|cc}
           & PSNR (dB) & SSIM (\%) \\
     \midrule
    UNet         & 24.21 & 90.85 \\
    XDT          & 24.66 & 91.32 \\
    \bottomrule
  \end{tabular}
  \label{tab:dissect}
\end{table}

\begin{figure*}[htp]
    \centering
    \includegraphics[width=0.8\textwidth]{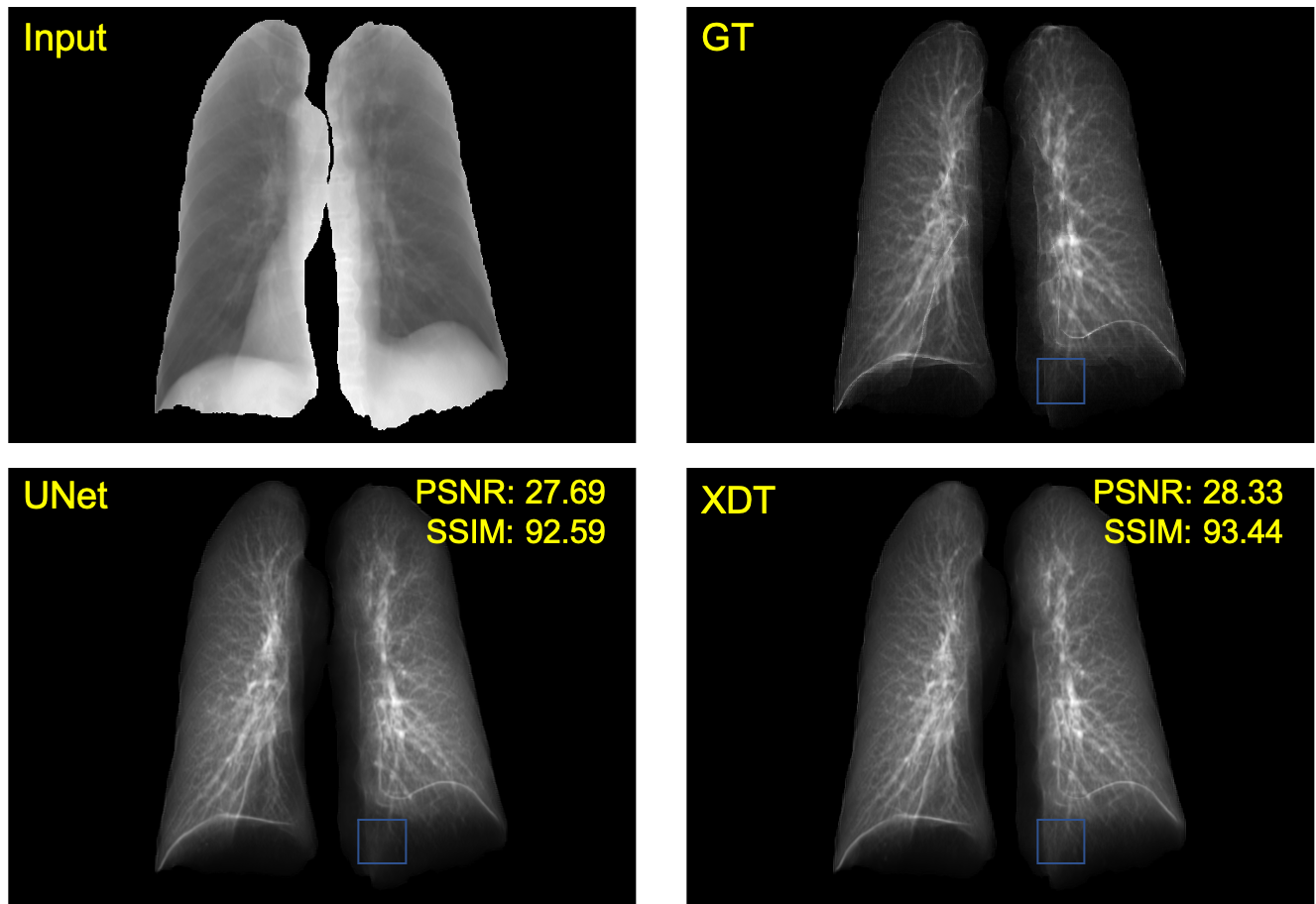}
    \caption{Dissection results using UNet and XDT respectively, where the comparison of the details in the blue rectangles shows the effectiveness of XDT.}
    \label{fig:dissect}
\end{figure*}

We randomly divided the dataset into a training set including 80\% patient data and a testing set including the remaining 20\% data.
Here projections were simulated along the three angles: -35$^{\circ}$, 0$^{\circ}$, and 35$^{\circ}$, where 0$^{\circ}$ is the front projection.
Before applying the XDT network, we trained a UNet model to segment lung field from the 2D projections so that the XDT network can focus on the lungs only.
We mainly evaluated the effectiveness of the XDT network. As show in Fig. \ref{fig:dissect}, the proposed XDT method faithfully dissected lungs and removed all other organs/tissues from 2D projections. The lung structures in the dissected image are much clearer than those in the original projections.
Furthermore, we compared the XDT network with the UNet baseline, where XDT network adds the back-projection and projection layers in the 5-th block in the baseline UNet.
During training, different projections were fed into the UNet model separately and into the XDT network collaboratively.
The quantitative results in Table \ref{tab:dissect} show that XDT network achieves slightly better results than the UNet baseline in terms of both PSNR and SSIM values.
The visual results in Fig. \ref{fig:dissect} also show that the XDT network can extract more details as indicated by the blue rectangles,
which demonstrate the effectiveness of the XDT framework.
By increasing the number of projection views, the performance of the XDT network could be further improved in principle, as more information is integrated through the back-projection and projection layer.

\subsection{Detection Results}

\begin{figure*}[htp]
    \centering
    \includegraphics[width=1\textwidth]{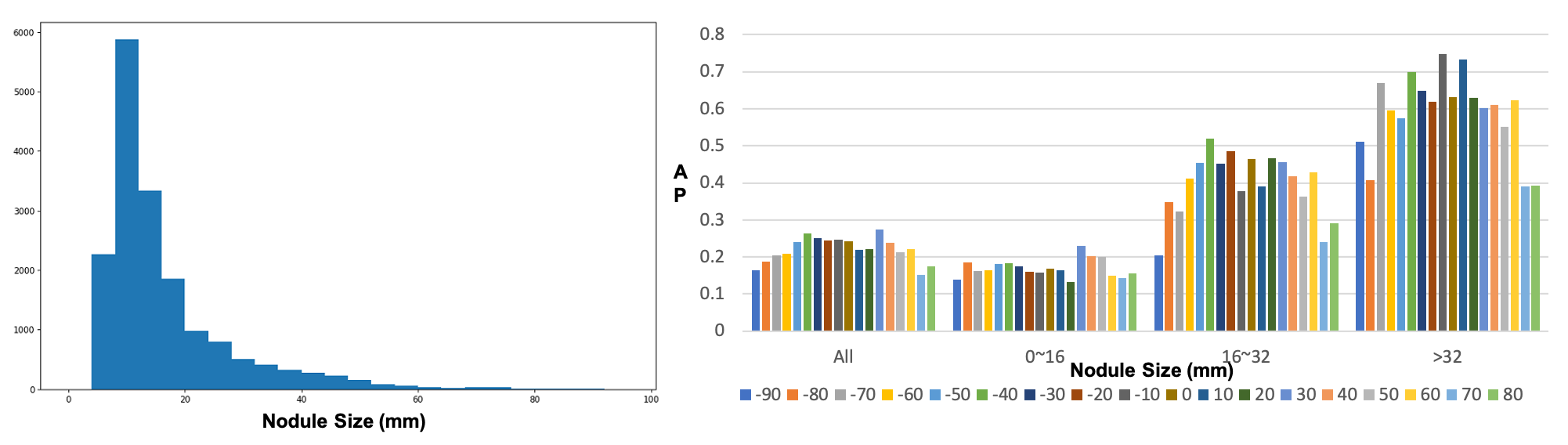}
    \caption{Nodule size distribution and detection performance over different projection angles.}
    \label{fig:views}
\end{figure*}

As mentioned in the Introduction section, the tumor visibility is significantly affected by the projection view angle. To this end, we first analyzed the nodule detection performance over different projection angles. In this experiment, all nodules $>$ 3mm accepted by at least three out of the four radiologists were included, and 18 projection views at [-90$^{\circ}$, -80$^{\circ}$, ..., 70$^{\circ}$, 80$^{\circ}$] were generated. The original Faster RCNN \cite{ren2015faster} model was trained. The nodule size distribution and performance distribution are shown in Fig. \ref{fig:views}. It can be seen that most the annotated nodules based on the CT scans in the LIDC-IDRI dataset are small, and the detection performance of such small nodules from 2D projections are relatively low in terms of average precision (AP). On the other hand, the performance varied over different angles, and basically the detection performance near the lateral side is significantly lower than that for other views.
Typical screening usually takes the front and lateral views for diagnosis of lung diseases.
These two orthogonal views can help in 3D localization of the lung nodule if the nodule can be detected in both views.
However, the above results indicate that nodules in the lateral view are usually missed so that 3D positions cannot be determined.

\begin{table}[htp]
  \renewcommand{\arraystretch}{1.5}
  \renewcommand\tabcolsep{20pt}
 \caption{Quantitative detection results.}
  \centering
  \begin{tabular}{c|cccc}
       AP@0.1    & -35$^{\circ}$ & 0$^{\circ}$ & 35$^{\circ}$ & ALL \\
     \midrule
    Separate         & \underline{0.254} & \underline{0.415} & \underline{0.402} & \underline{0.347} \\
    Separate-Collaborative         & 0.482 & 0.559 & 0.566 & 0.529 \\
    Dissected-Separate         & 0.209 & 0.497 & 0.330 & 0.329 \\
    Dissected-Collaborative         & \textbf{0.642} & \textbf{0.606} & \textbf{0.627} & \textbf{0.621} \\
    GT-Dissected-Separate         & 0.611 & 0.809 & 0.685 & 0.694 \\
    GT-Dissected-Collaborative         & 0.785 & 0.842 & 0.812 & 0.812 \\
    \bottomrule
  \end{tabular}
  \label{tab:detect}
\end{table}

\begin{figure*}[htp]
    \centering
    \includegraphics[width=1\textwidth]{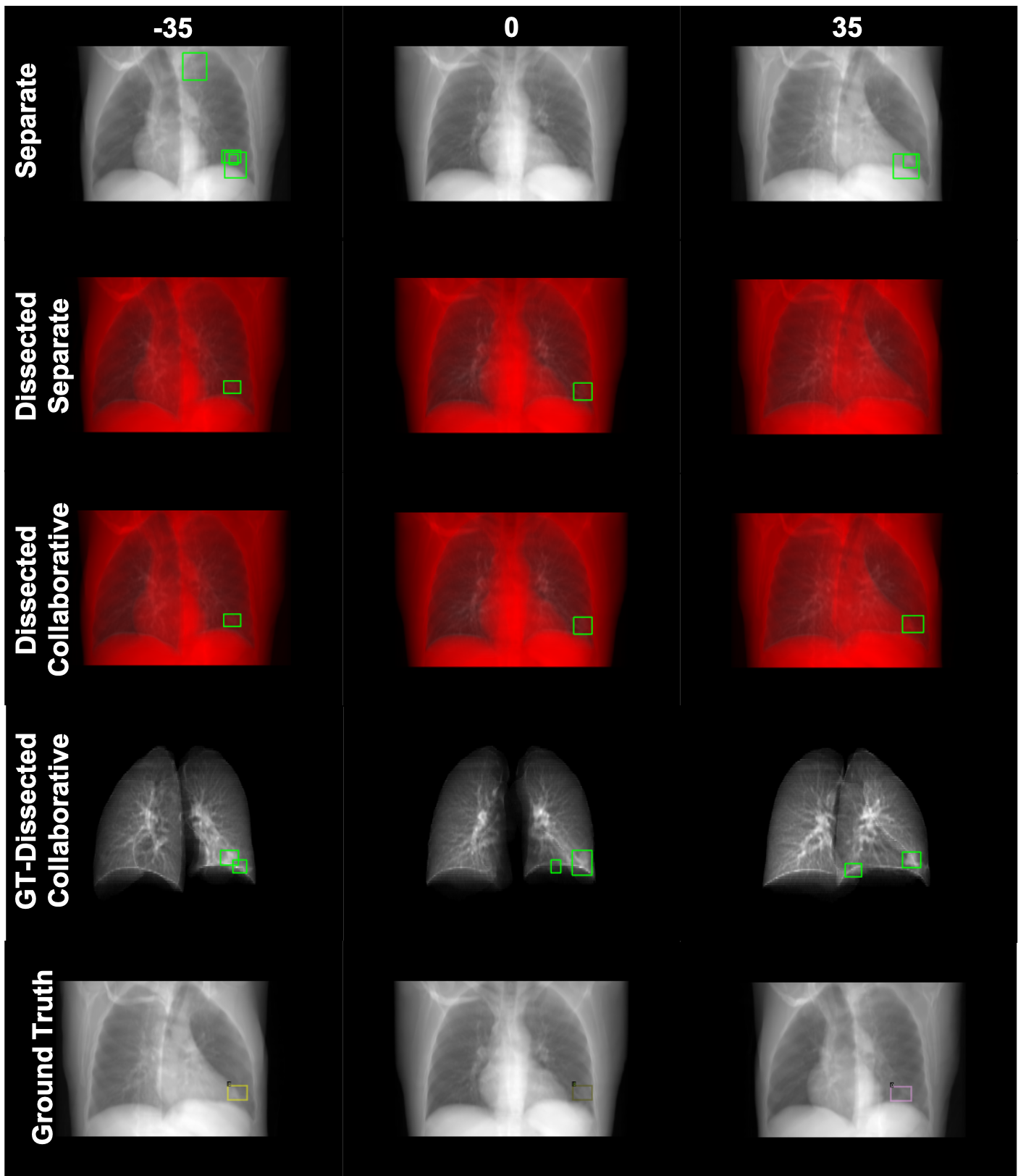}
    \caption{Lung nodule detection results using different methods. For Dissected-Separate and Dissected-Collaborative, the projection images consist of an original projection image in red channel and two dissected images in green and blue channels respectively.}
    \label{fig:results}
\end{figure*}

Based on the above results, we empirically selected three projection views: [-35$^{\circ}$, 0$^{\circ}$, 35$^{\circ}$] for training and testing the proposed collaborative detection network.
In this study, we only considered lung nodules $\ge$ 20mm accepted by at least three out of the four radiologists given the well-known difficulty of detecting nodules in 2D projections.
In this setting, 108 patient data were selected, 84 patients were randomly selected for training and the other 24 patients for testing. 
The XDT network was trained with 497 patient data including the 84 patients used for training the detection model but not including the 24 patients in the testing dataset.
The results in Table~\ref{tab:detect} show that the proposed detection method denoted by Dissected-Collaborative performed significantly better than the baseline method denoted by Separate, with the AP being improved by 20+\%.
The XDT dissection results improved the detection performance by about 10\% AP (Separate v.s. Dissected-Collaborative).
The 2D-3D matching algorithm also significantly improved the detection performance by 10+\% (Separate v.s. Separate-Collaborative, Dissected-Separate v.s. Dissected-Collaborative, and GT-Dissected-Separate v.s. GT-Dissected-Collaborative).
Also, if the lung dissection results are perfect, the proposed detection model can achieve 80+\% AP.
Typical visualization results in Fig. \ref{fig:results} show that the proposed method (Dissected-Collaborative) can accurately detect the lung nodules from all projection views, allowing accurate 3D localization.
All the above results demonstrate that our proposed XDT and collaborative detection networks are promising to improve the lung nodule detection performance on a very limited number of 2D X-ray projections; for example, three radiograms.

\section{Conclusion}
In conclusion, we have proposed the x-ray dissectography (XDT) and collaborative detection networks for detecting lung nodules from more than one 2D X-ray projections, and in particular only three planning radiographs.
The experimental results on the simulated dataset show that both the XDT and collaborative detection networks work jointly to improve the lung nodule detection performance significantly.
In the future, we plan to continue improving our networks and evaluating their effectiveness on real clinical datasets.
Hopefully, our results would help innovate the current x-ray radiographic imaging workflow from single or dual projections to triple radiographs or planning radiographs for satisfactory diagnostic information at minimal radiation dose in the cases of lung-related and other diseases.

\bibliography{scibib}

\bibliographystyle{Science}

\end{document}